\documentclass[prl,aps,twocolumn,superscriptaddress]{revtex4-1}

\usepackage{amsmath,amssymb,amsfonts,color,graphicx,tabularx}

\usepackage[unicode=true,colorlinks=true]{hyperref}

\hypersetup{linkcolor=blue,citecolor=blue,urlcolor=blue}

\begin{document}

\title{Tensor network representations of parton wave functions}

\author{Ying-Hai Wu}
\affiliation{School of Physics and Wuhan National High Magnetic Field Center, Huazhong University of Science and Technology, Wuhan 430074, China}

\author{Lei Wang}
\affiliation{Beijing National Lab for Condensed Matter Physics and Institute of Physics, Chinese Academy of Sciences, Beijing 100190, China}
\affiliation{Songshan Lake Materials Laboratory, Dongguan, Guangdong 523808, China}

\author{Hong-Hao Tu}
\email{hong-hao.tu@tu-dresden.de}
\affiliation{Institut f\"ur Theoretische Physik, Technische Universit\"at Dresden, 01062 Dresden, Germany}

\begin{abstract}
Tensor network states and parton wave functions are two pivotal methods for studying quantum many-body systems. This work connects these two subjects as we demonstrate that a variety of parton wave functions, such as projected Fermi sea and projected fermionic or bosonic paired states, can be represented exactly as tensor networks. The results can be compressed into matrix product states with moderate bond dimensions so various physical quantities can be computed efficiently. For the projected Fermi sea, we develop an excellent compression scheme with high fidelity using maximally localized Wannier orbitals. Numerical calculations on two parton wave functions demonstrate that our method exceeds commonly adopted Monte Carlo methods in some aspects. It produces energy and correlation function with very high accuracy that is difficult to achieve using Monte Carlo method. The entanglement measures that were almost impossible to compute before can also be obtained easily using our method.
\end{abstract}

\maketitle

{\em Introduction} --- The complexity of quantum many-body systems has posed considerable challenges for physicists since the dawn of quantum mechanics. One fundamental curse is that the Hilbert space of a composite system grows exponentially with the number of its constituents. While perturbative methods have been very successful in studying weak interactions, the vast arena of strongly correlated quantum matter remain elusive in many aspects. Analytical and numerical progresses have been made along various directions. The subjects of this Letter are tensor network states~\cite{verstraete2008,Cirac2009,schollwock2011,stoudenmire2012,orus2014,orus2018} and parton wave functions~\cite{anderson1987,arovas1988,Jain1989,WenXG1991b}, which share the common feature of trying to encode quantum many-body states using a moderate amount of resources.

Tensor network states are designed to capture special quantum entanglement patterns in the low-energy eigenstates of physical Hamiltonians. The wave functions are expressed as contraction of tensors (i.e., multi-index number arrays). If a system is divided into two subsystems, the entanglement entropy of one subsystem is bounded by the number of virtual indices on the boundary. In many cases, the number of parameters stays constant or grows polynomially, so the approximation is very useful. This approach begins with the invention of the density-matrix renormalization group (DMRG) algorithm~\cite{white1992} and has produced very impressive analytical and numerical results ever since.

The idea of parton wave functions was originally conceived in particle physics but has also been very successful in condensed matter physics. In this approach, the physical particles or spins are represented using slave particles (bosons or fermions) in certain enlarged Hilbert spaces. It is hoped that strongly correlated physical states can be approximated as suitable ``mean field" states of the slave particles with their unphysical components removed by some kind of projection. While this may appear to be {\it ad hoc} at first sight, it does provide very valuable insights into many problems. The ground states of some exactly solvable models, such as the Haldane-Shastry model~\cite{haldane1988,shastry1988} and the Kitaev honeycomb model~\cite{kitaev2006b}, can be expressed as Gutzwiller projected parton states. In the studies of high-$T_{c}$ superconductors~\cite{anderson2004,LeePA2006,edegger2007}, fractional quantum Hall states~\cite{McGreevy2012,LuYM2012,WuYH2017,KimY2019}, and quantum spin liquids~\cite{WenXG2002,Savary2016,zhou2017}, parton wave functions have been used extensively as variational ansatz.

It is usually possible to deduce some properties of parton wave functions using low-energy effective field theories~\cite{LeePA2006,zhou2017}. Nevertheless, numerical results are very much desired for quantitative assessments. For example, finding the optimal parameters with respect to a given Hamiltonian requires energy minimization. Monte Carlo methods are widely used for computing expectation values~\cite{RanY2007,Grover2010,Tay2011,Iqbal2011,Nataf2016,HuWJ2016}. This is relatively simple if the target state is made of fermionic determinants and/or Pfaffians but rather challenging if bosonic permanents are involved. The computation of entanglement entropy and entanglement spectrum~\cite{kitaev2006a,levin2006,LiH2008,QiXL2012,Dubail2012}, which have been used extensively to characterize many-body states, is still quite demanding for generic parton wave functions~\cite{ZhangY2011a,PeiJ2013,LiuZX2014,ShaoJ2015,Wildeboer2017}.

In this Letter, we prove that generic parton wave functions can be expressed as local tensor networks in a straightforward manner. The explicit representations of projected Fermi sea and projected fermionic or bosonic paired states correspond to sequential operations of matrix product operators (MPO) on simple product states. These tensor networks can be compressed into matrix product states (MPS) and various physical quantities can be evaluated efficiently. For the project Fermi sea, an optimized basis transformation using maximally localized Wannier orbitals is proposed, which greatly reduces the amount of entanglement in intermediate steps and helps to achieve high fidelity compressions. One can reach very high precision when computing physical quantities and directly access certain measures of quantum entanglement using the tensor network representations of parton wave functions. The numerical results clearly suggest that our method has the potential to surpass conventional Monte Carlo methods in many cases.

\begin{figure}
\includegraphics[width=0.48\textwidth]{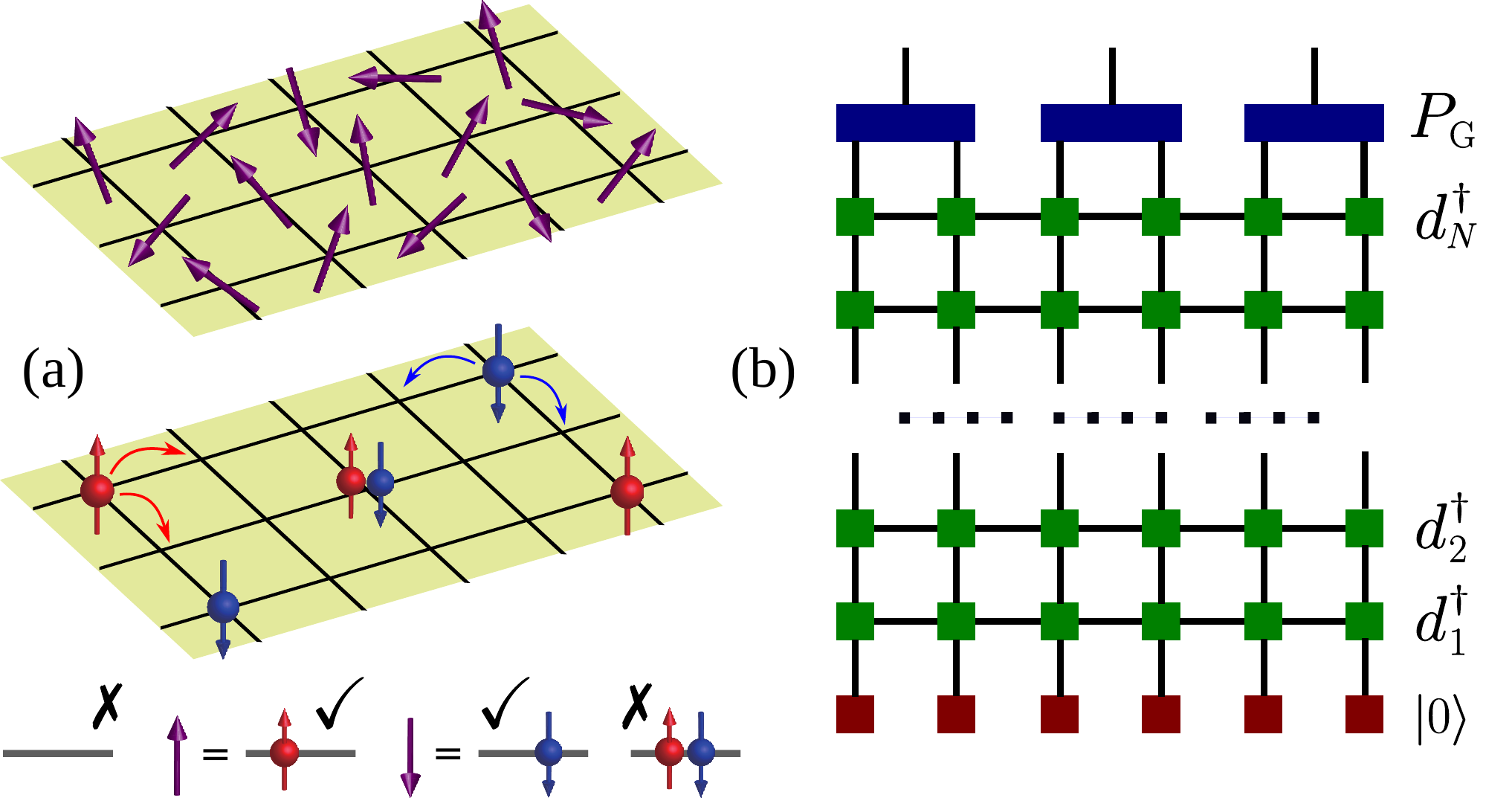}
\caption{(a) Schematics of parton construction for spin-1/2 lattice models. (b) Schematics of the tensor network representation of the projected Fermi sea in Eq.~(\ref{eq:ProjFS}).}
\label{Figure1}
\end{figure}

{\em Tensor network representation} --- The method proposed here can be applied to any spin, bosonic, or fermionic systems, but we shall use spin-1/2 lattice models as illustrations [see Fig.~\ref{Figure1}(a)]. The lattice sites are labeled by $j\in[1,N]$ and the spin operators are $S^{a}_{j}$ $(a=x,y,z)$. The Abrikosov fermion representation is $S^{a}_{j} = \frac{1}{2} \sum_{\alpha\beta} c^{\dag}_{j\alpha} \tau^{a}_{\alpha\beta} c_{j\beta}$, where $c^{\dag}_{j\alpha}$ ($c_{j\alpha}$) are fermionic creation (annihilation) operators at $j$, $\alpha=\uparrow,\downarrow$ is the spin index, and $\tau^a$ are Pauli matrices. This is an overcomplete representation with unphysical states (empty and doubly occupied) that need to be removed by the single-occupancy constraint $\sum_{\alpha} c^{\dag}_{j\alpha} c_{j\alpha} = 1$. The Schwinger boson representation is very similar, where the fermionic operators are replaced by their bosonic counterparts.

One popular class of trial wave functions for spin models is the projected Fermi sea
\begin{equation}
|\Psi\rangle = P_{\rm G} \prod^{N}_{m=1} d^{\dag}_{m} |0\rangle,
\label{eq:ProjFS}
\end{equation}
where $|0\rangle$ is the vacuum, the $d^\dag_{m}$'s are single-particle orbitals of the partons, $P_{\rm G}=\prod^{N}_{j=1} P_{j}$ is a product of projectors that impose the single-occupancy constraints on each site. In general, the single-particle orbitals can be written as $d^{\dag}_{m} = \sum^{N}_{j=1} \sum_{\alpha=\uparrow,\downarrow} A_{m,j\alpha} c^{\dag}_{j\alpha} = \sum^{2N}_{l=1} A_{ml} c^{\dag}_{l}$ with $l=(j,\alpha)$. The states labeled by $l$ are placed on a one-dimensional chain under some physically motivated guidelines~\cite{appendix}. This is in sharp contrast to previous works that construct (possibly nonlocal) tensor networks for parton wave functions~\cite{gu2010} or their norms~\cite{beri2011} on the original lattice. The $N{\times}2N$ matrix $A_{ml}$ that parametrizes the occupied orbitals is usually obtained by solving some ``mean-field'' Hamiltonians that are quadratic in the parton operators.

The central result of this Letter is that Eq.~(\ref{eq:ProjFS}) has a very natural tensor network representation. More importantly, it can be compressed into MPS with moderate bond dimensions, which allows for efficient computation of variational energy, correlation functions, and entanglement measures. The key observation that leads to our result is that the single-particle orbital $d^{\dag}_{m}$ can be converted to an MPO with bond dimension $D=2$ as~\cite{appendix}
\begin{eqnarray}
d^{\dag}_{m} =
\begin{pmatrix}
0 & 1
\end{pmatrix}
\left[
\prod^{2N}_{l=1}
\begin{pmatrix}
1 & 0 \\
A_{ml} c^{\dag}_{l} & 1
\end{pmatrix}
\right]
\begin{pmatrix}
1 \\
0
\end{pmatrix}
.
\label{eq:orbitalMPO1}
\end{eqnarray}
One dummy column and one dummy row are appended to ensure that all MPOs in the product have the same form. If the dummy vectors are multiplied with their neighbors, we recover a usual MPO with open boundary condition. It is then straightforward to find the tensor network representation of Eq.~(\ref{eq:ProjFS}) as depicted in Fig.~\ref{Figure1} (b): 1) apply the $N$ MPOs corresponding to the $d^{\dag}_m$'s to the fermionic vacuum; 2) apply the projector $P_{\rm G}$ to the Fermi sea with each term $P_j$ acting on two neighboring sites. In the same spirit, tensor network representations of projected fermionic or bosonic paired states can be obtained using MPOs that create fermionic or bosonic pairs~\cite{appendix,JinHK2020}.

{\em Compressing into MPS} --- Although the representation derived above is {\em exact}, physical quantities cannot be computed simply. In fact, it is well known that the exact contraction of a two-dimensional tensor network with closed loops is exponentially difficult~\cite{Cirac2009,orus2014}. This makes it imperative to develop an approximation scheme that would enable actual calculations. An obvious choice is to sequentially act the MPOs on the MPS (with fermionic vacuum as the initial input) to generate another MPS. However, the bond dimension of the MPS increases exponentially with the number of MPOs, so it is impossible to carry out the procedure for more than $\sim 12$ MPOs. To this end, we need to truncate the MPS at intermediate steps such that its bond dimension $D$ never exceeds some fixed values. The simplest truncation method is the singular value decomposition, where one converts the MPS into the so-called mixed canonical form and discards small singular values~\cite{schollwock2011,appendix}. Its efficiency is determined by the entanglement properties of the target state and its error is quantified by the norm of the discarded singular values.

If $A_{ml}$ in $d^{\dag}_{m}$ have similar magnitudes, it would substantially modify the matrices on all lattice sites when acting on an MPS, then the truncation is likely to introduce considerable errors. This is often the case since $d^\dag_m$ are eigenmodes of parton ``mean-field'' Hamiltonians, where $A_{ml}$ describe spatially extended Bloch waves or standing waves. It has been found that the single-particle orbitals and the sequence of applying MPOs can be optimized to minimize entanglement growth~\cite{Legeza2003-1,Legeza2003-2,Murg2010,Murg2012,Krumnow2016,Pastori2019}. In our case, the maximally localized Wannier orbitals~\cite{Wannier1937,Kohn1959,kivelson1982,qi2011b,Marzari2012} are adopted to facilitate the truncation. The basic idea is to convert the wave function in Eq.~(\ref{eq:ProjFS}) to $|\Psi\rangle = P_{\rm G} \prod^{N}_{r=1} \zeta^{\dag}_{r} |0\rangle$, where $\zeta^{\dag}_{r}$ are linear combinations of $d^{\dag}_{m}$. The entanglement entropy grows much slower when using the MPOs built from $\zeta^{\dag}_{r}$ because each one of them only causes appreciable changes (i.e., entanglement increase) in the vicinity of a particular lattice site. This is possible if $\zeta^{\dag}_{r}$ are designed to mimic the maximally localized Wannier orbitals. To be specific, the position operator $X=\sum^{N}_{j=1} \sum_{\alpha=\uparrow,\downarrow} jc^{\dag}_{j\alpha} c_{j\alpha}$ is expressed as a matrix~\footnote{It is assumed that $d^\dag_m$ have been orthogonalized to form a set of orthonormal modes with $\{d_m,d^\dag_n\}=\delta_{mn}$.}
\begin{eqnarray}
\widetilde{X}_{mn} = \langle 0 | d_{m} X d^{\dag}_{n} | 0\rangle
\end{eqnarray}
in the subspace spanned by $d^{\dag}_{m}$. Its eigenvectors are denoted using a matrix $B$ such that $B^{\dag}\widetilde{X}B$ is diagonal. The transformed orbital $\zeta^{\dag}_{r}$ is defined using $B_{mr}$ as
\begin{eqnarray}
\zeta^{\dag}_{r} = \sum^{N}_{m=1} B_{mr} d^{\dag}_{m} = \sum^{2N}_{l=1} (B^T A)_{rl} c^{\dag}_{l}.
\end{eqnarray}
The parton wave function is unchanged because $\zeta^{\dag}_{r}$ are just linear combinations of the same set of orbitals. In many cases, $d^{\dag}_{m}$ do not mix partons with different spins, so they can be separated to two groups that are transformed using the spin-up and spin-down position operators, respectively. As the order of $\zeta^{\dag}_{r}$ in $|\Psi\rangle$ does not matter, the truncation error can be further reduced by a ``left-meet-right" strategy: alternately act the operator localized at the left or right edge and gradually move toward the center.

\begin{table}
\begin{tabular}{ccc}
\hline
\hline
$D$    & energy & deviation \\
\hline
$1000$ \quad & \quad $-41.14354115$ \quad & \quad $3.7{\times}10^{-4}$ \\
$2000$ \quad & \quad $-41.14387956$ \quad & \quad $3.4{\times}10^{-5}$ \\
$3000$ \quad & \quad $-41.14390614$ \quad & \quad $7.2{\times}10^{-6}$ \\
$4000$ \quad & \quad $-41.14391112$ \quad & \quad $2.2{\times}10^{-6}$ \\
$5000$ \quad & \quad $-41.14391248$ \quad & \quad $8.6{\times}10^{-7}$ \\
\hline
\hline
\end{tabular}
\caption{Energy of the MPO-MPS results for the Haldane-Shastry model with $N=100$ at several different bond dimensions. The deviation is computed with respect to the exact ground state energy $-41.14391334$.}
\label{Table1}
\end{table}

\begin{figure}
\includegraphics[width=0.48\textwidth]{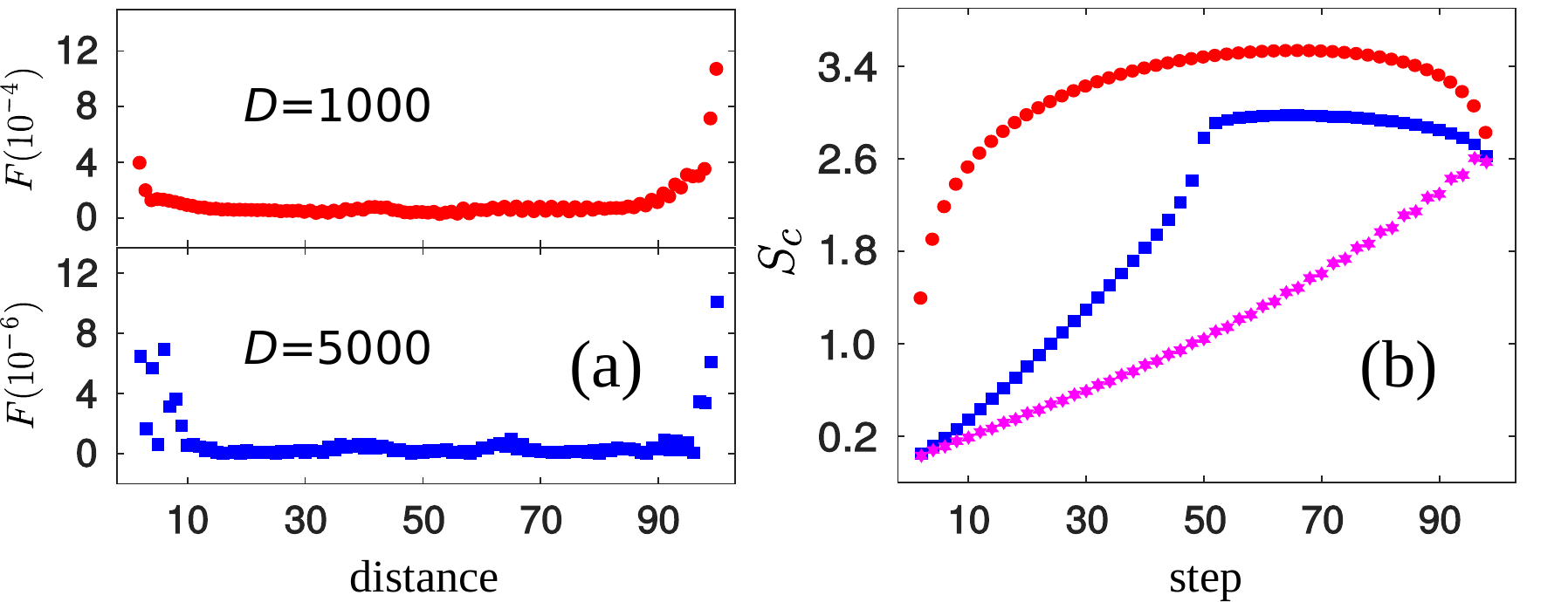}
\caption{(a) The absolute difference $F$ between the numerical and exact values of the spin-spin correlation function in the $N=100$ system. (b) The evolution of the von Neumann entanglement entropy $S_{c}$ at the center of the $N=100$ system during the calculation. Three methods are compared: (1) the original modes in $|\Psi_{\rm HS} \rangle$ (red dots), (2) the Wannier transformed modes from left to right (blue squares), and (3) the Wannier transformed modes and the left-meet-right strategy (magenta hexagons).}
\label{Figure2}
\end{figure}

{\em Numerical results 1} --- The first example that we have investigated is the Haldane-Shastry model~\cite{haldane1988,shastry1988} with the Hamiltonian
\begin{eqnarray}
H_{\rm HS} = \sum_{p<q} \frac{\pi^{2} \; {\mathbf S}_{p} \cdot {\mathbf S}_{q}}{N^{2}\sin^{2}\frac{\pi}{N}(p-q)} .
\label{eq:HS-PBC}
\end{eqnarray}
Its ground state for even $N$ is a Gutzwiller projected half-filled Fermi sea $|\Psi_{\rm HS} \rangle = P_{\rm G} \prod_{m} \prod_{\alpha=\uparrow,\downarrow} d^{\dag}_{m\alpha} |0\rangle$, where $d^{\dag}_{m\alpha}=N^{-1/2}\sum^{N}_{j=1} e^{-i(jm)} c^{\dag}_{j\alpha}$ is the creation operator in momentum space and the occupied momenta are $m=\frac{2\pi}{N}s$ with
\begin{eqnarray}
s=\left\{
\begin{array}{c}
0,{\pm}1,\ldots,\pm(\frac{N}{4}-1),\frac{N}{4} \\
0,{\pm}1,\ldots,\pm\frac{N-2}{4}
\end{array}
\right.\left.
\begin{array}{c}
\text{if } N \; {\rm mod} \; 4 = 0 \\
\text{if } N \; {\rm mod} \; 4 = 2
\end{array}
\right. .
\end{eqnarray}
The ground-state energy is $-\pi^{2}(N+5N^{-1})/24$ and the spin-spin correlation function in the ground state is~\cite{kuramotoBook,nielsen2011}
\begin{eqnarray}
\langle {\mathbf S}_{p} \cdot {\mathbf S}_{p+q} \rangle = \frac{ \sum^{N/2}_{a=1} \frac{3(-1)^{q}}{2a-1} \sin \left[ \frac{\pi}{N} (2a-1) q \right] }{ 2N \sin\frac{\pi}{N}q}.
\end{eqnarray}

The Haldane-Shastry parton state with $N=100$ has been constructed using our method for bond dimension $D$ up to $5000$. The comparison between the energy values in Table~\ref{Table1} and the spin-spin correlation function in Fig.~\ref{Figure2}(a) clearly demonstrates the success of our method. This level of accuracy is very difficult to achieve using Monte Carlo methods~\cite{edegger2007,zhou2017}. The evolution of the von Neumann entanglement entropy at the center of the system during the calculation is presented in Fig.~\ref{Figure2}(b). It is apparent that the Wannier mode transformation and the left-meet-right strategy are both very useful as they significantly reduce the amount of entanglement. The Haldane-Shastry model is difficult to study using direct DMRG method due to its gapless nature and the long-range interaction.

\begin{figure}
\includegraphics[width=0.48\textwidth]{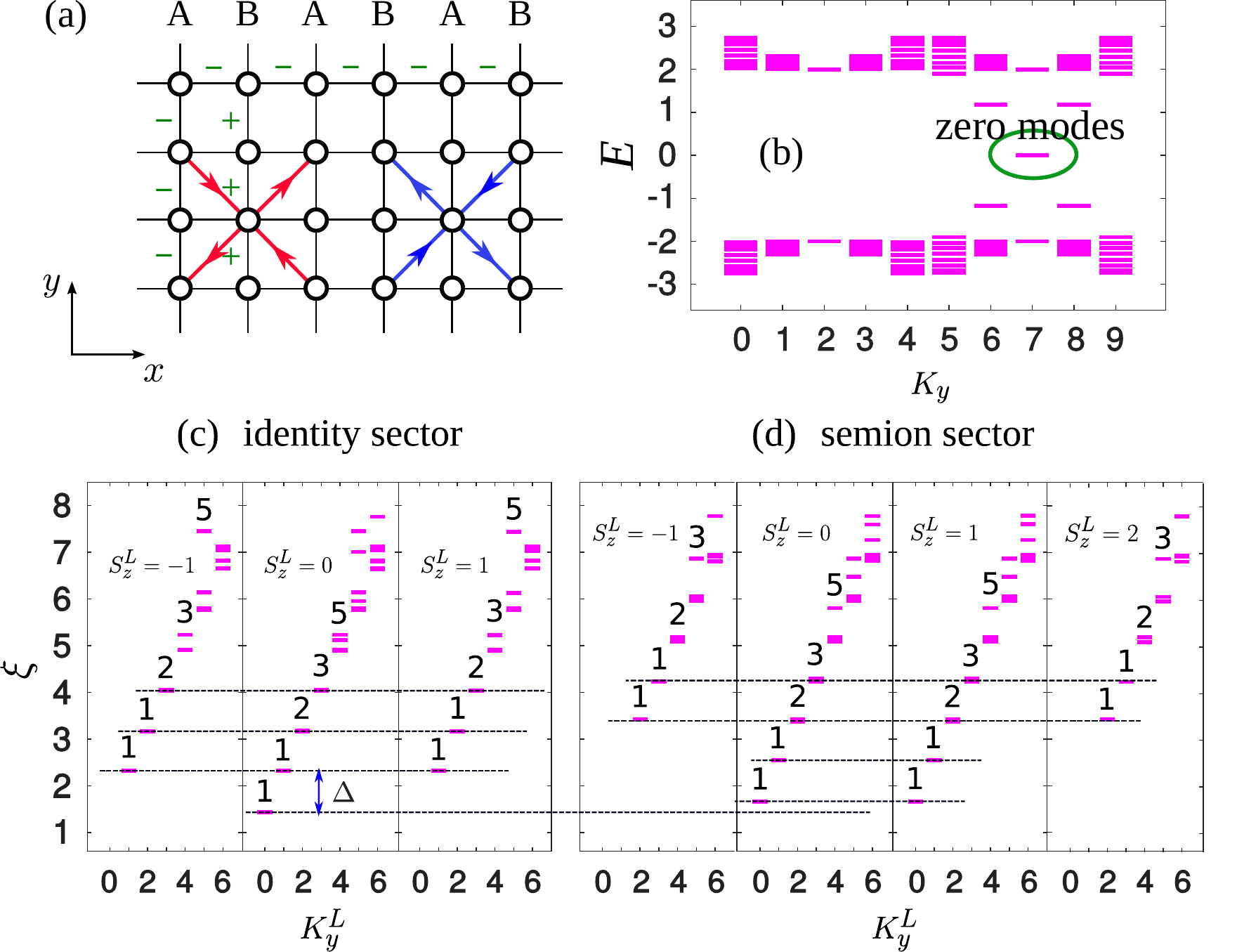}
\caption{(a) Schematics of the parton Hamiltonian of the chiral spin liquid model. Each unit cell contains two lattice sites labeled as $A$ and $B$. The signs of $t_{ij}$ are indicated using $\pm$ along the bonds. The signs of $\Delta_{jk}$ are negative (positive) along (against) the arrows on the colored lines. (b) The parton energy spectrum of the system with $N_{x}=16$ and $N_{y}=10$ on the cylinder with $\Theta_{y}=\pi$. There are two exact zero modes $d^{\dag}_{L\alpha}$ and $d^{\dag}_{R\alpha}$ for each spin that are localized at the left and right edges. (c,d) The entanglement spectrum of $|\Psi_{1}\rangle$ and $|\Psi_{2}\rangle$. The dashed lines indicate two sets of conformal towers in the two panels.}
\label{Figure3}
\end{figure}

{\em Numerical results 2} --- The second example that we have investigated is a chiral spin liquid model~\cite{ZhangY2011a} that has the same topological order as the $\nu=1/2$ Laughlin quantum Hall state~\cite{Laughlin1983,kalmeyer1987}. It is defined on a square lattice with $N_{x}$ and $N_{y}$ sites along the two directions. The spin-up and spin-down partons are described by the Hamiltonian
\begin{eqnarray}
H_{\rm CI} = \sum_{\langle{jk}\rangle,\alpha} t_{jk} c^{\dag}_{j\alpha} c_{k\alpha} +  \sum_{\langle\langle{jk}\rangle\rangle,\alpha} i \Delta_{jk} c^{\dag}_{j\alpha} c_{k\alpha},
\end{eqnarray}
where $\langle{jk}\rangle$ ($\langle\langle{jk}\rangle\rangle$) indicates nearest (next-nearest) neighbors. The hopping amplitudes satisfy $|t_{jk}|=1.0$ and $|\Delta_{jk}|=0.5$ and their signs are given in Fig.~\ref{Figure3}(a). The partons can be used to generate the chiral spin liquid and we aim to compute its entanglement spectrum. The system is divided into two parts, the reduced density matrix of the left half is computed, and the entanglement spectrum (i.e., the negative logarithm of the eigenvalues of the reduced density matrix) is plotted versus the good quantum numbers. This is almost impossible to do for generic parton wave functions using current Monte Carlo methods and unambiguously demonstrates the power of our method.

It is preferable to consider the cylinder rather than the torus for our purpose~\cite{LiuZ2012}. The $y$ direction is chosen to be periodic and the associated boundary twist angle is $\Theta_{y}$. An important step is to find the minimally entangled states (MES) because topological information can be extracted most efficiently using them~\cite{Keski1993,zhang2012,Cincio2013,Zaletel2013,tu2013b,ZhuW2015}. This can be done if $N_{y}$ is a multiple of two (but not four) and $\Theta_{y}=\pi$ or if $N_{y}$ is a multiple of four and $\Theta_{y}=0$. For such systems, the energy spectrum of $H_{\rm CI}$ contains four exact zero modes $d^{\dag}_{L\alpha}$ and $d^{\dag}_{R\alpha}$ [see Fig.~\ref{Figure3}(b)], which reside at the left and right edges. The many-body state in which the negative energy single-particle orbitals are fully populated is denoted as $|\Phi\rangle$. The zero modes can be occupied in four different ways to generate
\begin{eqnarray}
&& |\Psi_{1}\rangle = P_{\rm G} d^{\dag}_{L\uparrow} d^{\dag}_{L\downarrow} |\Phi\rangle, \quad |\Psi_{2}\rangle = P_{\rm G} d^{\dag}_{L\uparrow} d^{\dag}_{R\downarrow} |\Phi\rangle, \nonumber \\
&& |\Psi_{3}\rangle = P_{\rm G} d^{\dag}_{L\downarrow} d^{\dag}_{R\uparrow} |\Phi\rangle, \quad |\Psi_{4}\rangle = P_{\rm G} d^{\dag}_{R\uparrow} d^{\dag}_{R\downarrow} |\Phi\rangle.
\end{eqnarray}
The numerical results quoted below are from the case with $N_{x}=16$ and $N_{y}=10$, but smaller systems have also been checked and the results are consistent. The only appreciable overlap between them at $D=8000$ is $|\langle\Psi_{1}|\Psi_{4}\rangle|=0.9237$ (others are smaller than $10^{-8}$), which agrees with the previous claim that $|\Psi_{1}\rangle=|\Psi_{4}\rangle$~\cite{tu2013b}. This means that there are three rather than two linearly independent states, so the choice of MES is a subtle issue, but it turns out that either $|\Psi_{1}\rangle$ and $|\Psi_{2}\rangle$ (or $|\Psi_{1}\rangle$ and $|\Psi_{3}\rangle$) can be used as the two MESs~\cite{appendix}. The entanglement spectra of $|\Psi_{1}\rangle$ and $|\Psi_{2}\rangle$ at $D=9000$ are shown in Figs.~\ref{Figure3}(c) and (d). The accuracy of these states is quantified by the many-body momentum $K_{y}$. The ideal expectation value of $\exp[iK_{y}N_{y}/(2\pi)]$ is $1$, whereas the numerical value is $0.9714$ for $|\Psi_{1}\rangle$ and $0.9955$ for $|\Psi_{2}\rangle$. The good quantum numbers for the entanglement levels are the $z$-component spin $S^{L}_{z}$ and the momentum $K^{L}_{y}$ of the left half. The characteristic chiral boson counting $1,1,2,3,5,\ldots$ are observed in all cases. The lowest entanglement eigenvalue of $|\Psi_{1}\rangle$ is smaller than that of $|\Psi_{2}\rangle$, so the former is the identity sector and the latter is the semion sector. The total countings agree with those of the SU(2)$_1$ Wess-Zumino-Witten model: $1,3,4,7,\cdots$ in the identity sector and $2,2,6,8,\cdots$ in the semion sector~\cite{francesco1997}. The topological spin $h$ of the semion can be computed as $h=(\xi_{0,{\rm s}}-\xi_{0,{\rm I}})/\Delta$, where $\xi_{0,{\rm I}}$ ($\xi_{0,{\rm s}}$) is the lowest entanglement eigenvalue in the identity (semion) sector and $\Delta$ is the spacing between the first two entanglement levels in the identity sector [see Fig.~\ref{Figure3}(c)]. Its numerical value $0.2617$ is reasonably close to the theoretical value $1/4$.

The identification of MES here reveals an important general feature about the entanglement structure of chiral topological phases enriched by a global symmetry (denoted by $G$). The topological sectors on the cylinder are labeled by definite anyon flux threading the cylinder. While the ground state is invariant under $G$, the two anyons may transform under a nontrivial (higher-dimensional and possibly projective) representation of $G$ and possess nonlocal entanglement. If one would like to obtain an entanglement spectrum that corresponds to a \emph{single} CFT tower (labeled by a primary associated with the anyon), the symmetry should be broken such that the anyons are projected onto certain ``product state'' and the nonlocal entanglement is destroyed. For the semion sector of the chiral spin liquid, we get a singlet state $|\tilde{\Psi}_{s}\rangle = |\Psi_{2}\rangle - |\Psi_{3}\rangle$ when the semions carrying spin-1/2 at the edges form a nonlocal singlet~\cite{appendix}. However, the MES in the semion sector should be taken as $|\Psi_{2}\rangle$ or $|\Psi_{3}\rangle$, where the semions at the edges are ``polarized''. This observation is very useful for studying chiral topological order using entanglement spectrum, especially when symmetries are implemented in DMRG simulations.

{\em Conclusion and discussion} --- In summary, we have constructed exact tensor network representations for generic parton wave functions. The tensor network representations take the form of sequential operations of MPO on simple product states and can be conveniently compressed into MPS. This allows one to characterize parton wave functions using powerful MPS techniques and greatly expands the utility of parton wave functions as variational ansatz. The parton wave functions studied in this Letter have no free parameters. An immediate next step is to consider some cases with variational parameters and search for their optimal values. The tensor network automatic differentiation method is well-adapted for this purpose~\cite{LiaoHJ2019,Hubig2019,WanZQ2019}. The parton wave functions could be supplied as initial inputs to speedup DMRG simulations~\cite{stoudenmire2012,Fishman2015}. Besides the ground states, our method is also capable of studying excitations. The numerical prospect of parton wave functions in the age of tensor networks deserves further investigations and we hope to report other interesting results in future works.

{\em Acknowledgment} --- We are grateful to Jan Carl Budich, Jan von Delft, Geza Giedke, Seung-Sup Lee, David Luitz, and Rom\'an Or\'us for stimulating discussions. This work was supported by the NNSF of China under Grant No. 11804107 and startup grant of HUST (YHW),  Ministry of Science and Technology of China under the Grant No.2016YFA0302400 (LW), and the DFG through project A06 (HHT) of SFB 1143 (Project-id 247310070).

\bibliography{Parton}

\clearpage

\setcounter{figure}{0}
\setcounter{equation}{0}
\renewcommand\thefigure{A\arabic{figure}}
\renewcommand\theequation{A\arabic{equation}}

\begin{widetext}

\section{Appendix A: Further technical details}

This section provides more technical details that are helpful in practical calculations. The fermionic creation and annihilation operators are not convenient to handle in tensor networks due to their anticommutation relation. To this end, the fermionic orbitals are converted to spin-1/2 degrees of freedom $\sigma_{l}$ (not to be confused with the original spin-1/2's) using the Jordan-Wigner transformation $c^{\dag}_{l} =\sigma^{z}_{1} \cdots \sigma^{z}_{l-1} \sigma^{+}_{l}$. The MPO for $d^{\dag}_{m}$ becomes
\begin{eqnarray}
d^{\dag}_m =
\begin{pmatrix}
0 & 1
\end{pmatrix}
\left[
\prod^{2N}_{l=1}
\begin{pmatrix}
1 & 0 \\
A_{ml} \sigma^{+}_{l} & \sigma^{z}_l
\end{pmatrix}
\right]
\begin{pmatrix}
1 \\
0
\end{pmatrix},
\label{eq:orbitalMPO2}
\end{eqnarray}
and the fermionic vacuum changes to $|\downarrow\downarrow\ldots\downarrow\rangle$ where all spins point down. In the projected Fermi sea state, the number of partons is a good quantum number and it translates to the total $z$-component spin after the Jordan-Wigner transformation. The vacuum has no parton and acting one $d^{\dag}_{m}$ increases the number of partons by $1$. For the two examples studied in the main text, the spin-up and spin-down modes are not mixed in any step, so the intermediate states have a U(1)$\times$U(1) symmetry corresponding to the numbers of partons with spin-up and spin-down. This symmetry can be exploited to significantly improve the computational speed.

The states labeled by $l$ can be placed on a one-dimensional chain in arbitrary orders, but some physical considerations suggest that they should be organized properly to simplify the computation. Two states on the same physical lattice sites but with opposie spins should be fixed as neighbors so the Gutzwiller projection on them can be performed easily. In principle, the Gutzwiller projection is implemented after the whole unprojected state has been generated. However, if only the ground state is concerned (as in the main text), all double occupancy in the intermediate steps can be removed immediately since they would not survive in the final projection. It is useful to merge two neighboring sites with spin-up and spin-down partons as a single site. The local Hilbert space dimension is $4$ if there is no constraint. One can simply discard the doubly occupied state and reduce the dimension to $3$. As for other tensor network methods, the MPO-MPS method favors parton wave functions with low entanglement. For parton wave functions in two dimensions, the cylinder geometry is more preferable than the torus geometry because quantum states often have larger entanglement on the latter. If each unit cell of the system contains more than one site (such as honeycomb and kagome lattices), it is not clear {\em a priori} how to choose the best ``one-dimensional path" such that the truncation error can be minimized when constructing the MPS representation. The experience accumulated during previous DMRG simulations could be very useful~\cite{stoudenmire2012}. It is also a good idea to perform benchmark calculations using different organizations of the one-dimensional path to compare the truncation errors. The entanglement structures of various types of parton wave functions are relatively well-understood (compared to the ground states of local Hamiltonians). For instance, gapped parton wave functions with topological order should satisfy the entanglement area law, while gapless parton states with a Fermi surface would mostly likely violate the entanglement area law and might be difficult to represent using MPS in large systems. Based on such prior knowledge, we could roughly estimate the computational resource required for generating a particular parton wave function using the MPO-MPS method.

The compression of MPS with large bond dimensions is an essential step in our method. In our calculations, the compression is performed using singular value decomposition (SVD) in a ``full-update" manner. In every MPO-MPS evolution step, a new MPS is generated with doubled bond dimension (as the MPO has bond dimension 2). The resulting MPS is brought to the left canonical form using QR decomposition (without any truncation), and then converted to the right canonical form with truncation using SVD on each bond. This means that the MPS is truncated in the mixed canonical form with the whole environment taken into account. The SVD-compressed result can be further improved if one uses it as input to perform some DMRG sweeps. This is very likely to improve the compression fidelity at the cost of longer computational time. In practice, we found that SVD with maximally localized Wannier orbitals already yields quite accurate results. Nevertheless, it could still be helpful to perform some DMRG sweeps when the entanglement is large, especially in the last few MPO-MPS steps.

\section{Appendix B: Projected fermionic or bosonic paired states}

Another important class of parton wave functions is projected fermionic or bosonic paired states. The bosonic paired state has the general form
\begin{equation}
|\Psi\rangle = P_{\rm G} \exp \left( \sum_{k{\neq}l} g_{kl} b^{\dag}_{k} b^{\dag}_{l} \right) |0\rangle,
\label{eq:ProjBCS1}
\end{equation}
where $k,l$ include both site and spin indices, $g_{kl}$ is the pairing function between them, and $b^{\dag}_{k}$ is the creation operator for the $k$-th bosonic mode. As for the projected Fermi sea in Eq.~(1) of the main text, the unprojected paired states in Eq.~(\ref{eq:ProjBCS1}) are usually obtained by solving some ``mean-field'' Hamiltonians of the partons with pairing terms.

The state in Eq.~(\ref{eq:ProjBCS1}) also has a natural tensor network representation. We again consider the spin-1/2 case for illustrating the method. Because of the single-occupancy constraint imposed by the Gutzwiller projector, Eq.~(\ref{eq:ProjBCS1}) can be rewritten as
\begin{eqnarray}
|\Psi\rangle = P_{\rm G} \prod^{2N}_{k=1} \prod^{2N}_{l=1({\neq}k)} \left( 1+g_{kl} b^{\dag}_{k} b^{\dag}_{l} \right) |0\rangle = P_{\rm G} \prod^{2N}_{k=1} W_{k} |0\rangle,
\label{eq:ProjBCS2}
\end{eqnarray}
where
\begin{eqnarray}
W_{k} = 1 + \sum^{2N}_{l=1({\neq}k)} g_{kl} b^{\dag}_{k} b^{\dag}_{l} =
\begin{pmatrix}
1 & 0
\end{pmatrix}
\left[
\prod^{k-1}_{l=1}
\begin{pmatrix}
1 & g_{kl} b^{\dag}_{l} \\
0 & 1
\end{pmatrix}
\right]
\begin{pmatrix}
1 & b^{\dag}_k \\
b^{\dag}_k & 0
\end{pmatrix}
\left[
\prod^{2N}_{l=k+1}
\begin{pmatrix}
1 & 0 \\
g_{kl} b^{\dag}_{l} & 1
\end{pmatrix}
\right]
\begin{pmatrix}
1 \\
0
\end{pmatrix}
\label{eq:WMPO1}
\end{eqnarray}
is explicitly expressed as an MPO with bond dimension $2$. The projected bosonic paired state can be obtained by successively applying $2N$ MPOs to the bosonic vacuum and performing Gutzwiller projection at the end. It is worth noting that the bosonic operators in Eq.~(\ref{eq:WMPO1}) actually create hardcore bosons due to the presence of $P_{\rm G}$. The role of $W_{k}$ is to create valence-bond singlets between site $k$ and other sites.

The bosonic partons in Eq.~(\ref{eq:ProjBCS1}) can be replaced by fermionic partons and the result can also be converted to a tensor network. For the fermionic paired state, we have the operator
\begin{eqnarray}
W_{k} &=&
\begin{pmatrix}
1 & 0
\end{pmatrix}
\left[
\prod^{k-1}_{l=1}
\begin{pmatrix}
1 & g_{kl} c^{\dag}_{l} \\
0 & 1
\end{pmatrix}
\right]
\begin{pmatrix}
1 & c^{\dag}_k \\
-c^{\dag}_k & 0
\end{pmatrix}
\left[
\prod^{2N}_{l=k+1}
\begin{pmatrix}
1 & 0 \\
g_{kl} c^{\dag}_{l} & 1
\end{pmatrix}
\right]
\begin{pmatrix}
1 \\
0
\end{pmatrix}
\label{eq:WMPO2}
\end{eqnarray}
for creating fermionic valence bonds. As for the projected Fermi sea, it is also convenient to perform a Jordan-Wigner transformation and use
\begin{eqnarray}
W_{k} &=&
\begin{pmatrix}
1 & 0
\end{pmatrix}
\left[
\prod^{k-1}_{l=1}
\begin{pmatrix}
1 & g_{kl} \sigma^{+}_l \\
0 & \sigma^{z}_l
\end{pmatrix}
\right]
\begin{pmatrix}
1 & -\sigma^{+}_k \\
\sigma^{+}_k & 0
\end{pmatrix}
\left[
\prod^{2N}_{l=k+1}
\begin{pmatrix}
1 & 0 \\
g_{kl} \sigma^{+}_l & \sigma^{z}_l
\end{pmatrix}
\right]
\begin{pmatrix}
1 \\
0
\end{pmatrix}.
\label{eq:WMPO}
\end{eqnarray}
An efficient compression scheme for projected fermionic paired states has been developed using maximally localized Wannier orbitals of Bogoliubov quasiparticles and quasiholes~\cite{JinHK2020}. It is still unclear whether a similar method can be devised for projected bosonic paired states.

\section{Appendix C: Minimally entangled states on the cylinder}

This section provides more details about the minimally entangled states (MESs) of the chiral spin liquid on the cylinder. For topologically ordered systems, the ground state is not unique on certain manifolds. This brings out the problem of choosing a suitable basis for multiple degenerate ground states. The MESs constitute an important basis because they minimize the entanglement entropy of non-contractible regions on the manifold~\cite{zhang2012}. In the present setup, the region is half of the cylinder.

As we have explained in the main text, filling the boundary zero modes in the parton energy spectrum results in three orthogonal states
\begin{eqnarray}
|\Psi_{1}\rangle = P_{\rm G} d^{\dag}_{L\uparrow} d^{\dag}_{L\downarrow} |\Phi\rangle, \quad |\Psi_{2}\rangle = P_{\rm G} d^{\dag}_{L\uparrow} d^{\dag}_{R\downarrow} |\Phi\rangle, \quad
|\Psi_{3}\rangle = P_{\rm G} d^{\dag}_{L\downarrow} d^{\dag}_{R\uparrow} |\Phi\rangle.
\end{eqnarray}
The orthogonality has been checked by computing their overlaps using standard MPS techniques. It is easy to see that $|\Psi_{1}\rangle$ is a spin-singlet. In contrast, $|\Psi_{2}\rangle$ and $|\Psi_{3}\rangle$ do not have definite total spins. One can use their linear combinations to form a spin-singlet
\begin{eqnarray}
|{\widetilde\Psi}_{s}\rangle = P_{\rm G} (d^{\dag}_{L\uparrow} d^{\dag}_{R\downarrow} - d^{\dag}_{L\downarrow} d^{\dag}_{R\uparrow}) |\Phi\rangle
\end{eqnarray}
and a spin triplet (within the $S_{z}=0$ subspace)
\begin{eqnarray}
|{\widetilde\Psi}_{t}\rangle = P_{\rm G} (d^{\dag}_{L\uparrow} d^{\dag}_{R\downarrow} + d^{\dag}_{L\downarrow} d^{\dag}_{R\uparrow}) |\Phi\rangle.
\end{eqnarray}
If the cylinder is wrapped to be a torus, $|\Psi_{1}\rangle$ and $|{\widetilde\Psi}_{s}\rangle$ would adiabatically evolve to the two degenerate ground states. The physical picture for $|{\widetilde\Psi}_{s}\rangle$ is that $d^{\dag}_{L\alpha}$ ($d^{\dag}_{R\alpha}$) creates a semion with spin projection $\alpha$ at the left (right) boundary, which ensures that $|{\widetilde\Psi}_{s}\rangle$ has a well-defined semion flux inside the cylinder.

For a system with chiral topological order, Ref.~\cite{QiXL2012} found that the reduced density matrix of an MES on the half-cylinder is a thermal state of a chiral conformal field theory (CFT)
\begin{eqnarray}
\rho_{a} \propto e^{-H_{\rm CFT}}|_{a}.
\end{eqnarray}
The subscript $a$ labels the anyon flux in the MES as well as the CFT primary field associated with the anyon. This result helps us to identify the MESs of the chiral spin liquid. As shown in Fig.~3(c) of the main text, the level counting in the entanglement spectrum of $|\Psi_{1}\rangle$ agrees with the identity sector of the SU(2)$_1$ Wess-Zumino-Witten (WZW) model, so we conclude that $|\Psi_{1}\rangle$ is the MES in the identity sector.

For the semion sector, some subtle issues arise when we try to find the MES. The entanglement spectrum of $|{\widetilde\Psi}_{s}\rangle$ does {\em not} exhibit the conformal towers of the spin-1/2 primary of the SU(2)$_1$ WZW model but turns out to be a tensor product of two identical copies of that. This is somewhat confusing at first sight but not really surprising. The two semions at the left and right edges in $|{\widetilde\Psi}_{s}\rangle$ have opposite spins and form a singlet. However, the MES in the semion sector should break this nonlocal entanglement by fixing the $S_z$ quantum number of the distant semions. This can only be achieved by mixing $|{\widetilde\Psi}_{s}\rangle$ and $|{\widetilde\Psi}_{t}\rangle$ properly to break down their SU(2) symmetry to U(1) symmetry. The MES in the semion sector can be chosen as $|\Psi_{2}\rangle$ or $|\Psi_{3}\rangle$. As shown in Fig.~3(d) of the main text, the level counting in the entanglement spectrum of $|\Psi_{2}\rangle$ agrees with the semion sector of the SU(2)$_1$ WZW model. This is also the case for $|\Psi_{3}\rangle$.

The situation encountered here is reminiscent of the spin-1 AKLT model on an open chain with finite length. This system has four degenerate ground states (one singlet and one triplet) due to two emergent spin-1/2 edge states. The entanglement spectrum of the singlet state on a half-chain has four quasi-degenerate levels, which is twice as much as the two-fold degeneracy expected for the spin-1 Haldane phase. To reveal the two-fold degeneracy, we need to project the two edge states to subspaces with fixed $S_{z}$ quantum numbers and then compute the entanglement spectrum. This can be done using a linear combination of the singlet and the triplet with $S_{z}=0$.

\section{Appendix D: Permanents from tensor network states}

\begin{figure}
\includegraphics[width=0.8\textwidth]{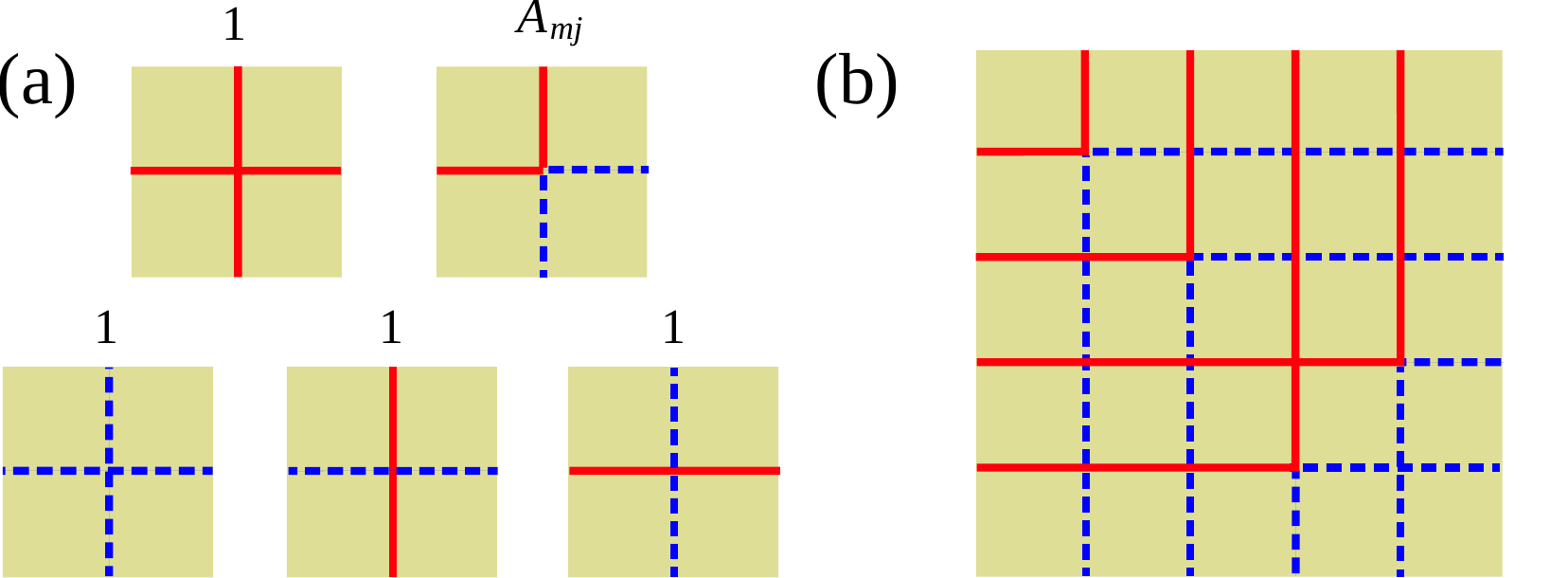}
\caption{(a) Nonzero elements in the tensor network representation of permanents. (b) One term in the tensor contraction of permanents, e.g., $A_{11}A_{22}A_{34}A_{43}$.}
\label{FigureA1}
\end{figure}

As a byproduct of our method, a tensor network representation of permanents can be designed. The permanent of a $N{\times}N$ matrix $A$ can be encoded using $N$ bosonic modes described by creation (annihilation) operators $b^{\dag}_{k}$ ($b_{k}$) with $k=1,2,\cdots,N$. Let us consider the many-body wave function
\begin{eqnarray}
|\Psi\rangle = d^{\dag}_{1} d^{\dag}_{2} \cdots d^{\dag}_{N} |0\rangle,
\label{eq:bosonperm}
\end{eqnarray}
where
\begin{eqnarray}
d^{\dag}_{m} = \sum^{N}_{j=1} A_{m,j} b^{\dag}_{j}, \qquad (1\leq m \leq N) .
\end{eqnarray}
The permanent ${\rm Per}(A)$ is the overlap
\begin{eqnarray}
\langle 0 | b_{N} b_{N-1} \cdots b_{1} |\Psi\rangle = \langle 0 | b_{N}b_{N-1} \cdots b_{1} \; d^{\dag}_{1} d^{\dag}_{2} \cdots d^{\dag}_{N} |0\rangle.
\label{eq:overlap}
\end{eqnarray}
It is apparent that the one-mode overlap $\langle 0| b_{j} d^{\dag}_{m} |0 \rangle = A_{m,j}$. The Wick's theorem tells us that the right hand side of Eq.~(\ref{eq:overlap}) can be expressed using the one-mode overlap as
\begin{eqnarray}
\sum_{p{\in}S_{N}} A_{1,p(1)}A_{2,p(2)} \cdots A_{N,p(N)}.
\end{eqnarray}
It consists of all possible combinations of $b_{j}$ and $d^{\dag}_{m}$ ($S_{N}$ is the permutation group of $N$ elements), which is precisely the definition of the permanent of $A$. The bosonic mode $d^{\dag}_{m}$ can be converted to an MPO as we have done for the fermionic mode in the main text. This helps us to find the tensor network representation of $|\Psi\rangle$. The overlap is the contraction of $|\Psi\rangle$ with the MPS $\langle 0| b_{N} b_{N-1} \cdots b_{1}$. The hardcore condition can be imposed on each site because the configurations with more than one boson on any site do not contribute to the permanent. This means that the physical legs of the MPO always have dimension $2$.

The tensor network representation of permanents has an appealing geometric picture that reveals its connection to the counting of perfect matching~\cite{moorebook2011}. As shown in Fig.~\ref{FigureA1}(a), the local tensor has only five nonzero elements, whose values are either $1$ or $A_{mj}$. The $m,j$ indices label the location of the tensor in the network. The binary indices are denoted using red solid and blue dashed lines, so they acquire geometric meaning of world lines. The tensor network for ${\rm Per}(A)=\langle 0| b_{N} b_{N-1} \cdots b_{1} |\Psi\rangle$ assembles local tensors into the form of Fig.~\ref{FigureA1}(b). One recognizes that the tensor contraction amounts to count the weighted sum of perfect matchings. In cases where all matrix elements are non-negative, there can be efficient stochastic algorithm to estimate the permanent by sampling the permutation of the world lines. The tensor network method can even deal with the cases where $A$ is a complex matrix. Moreover, suppose that one changes the tensor element of the first diagram in Fig.~\ref{FigureA1}(a) from $1$ to $-1$, the tensor contraction would then evaluate the matrix determinant instead of permanent. This is also intuitive from the picture since the tensor element corresponds to the crossing of the world lines.

After extensive numerical experiments, we conclude that this method is not as fast as the Ryser's algorithm with gray code. Nevertheless, we hope that this observation could be useful in some different settings, such as designing efficient algorithms for approximating permanents.

\end{widetext}

\end{document}